# Spectral Homogenization of the Radiative Transfer Equation via Low-Rank Tensor Train Decomposition

Y. Sungtaek Ju[a]

Department of Mechanical and Aerospace Engineering, University of California, 420 Westwood Plaza, Los Angeles, CA 90095-1597

**Abstract**

Radiative transfer in absorbing–scattering media requires solving a transport equation across a spectral domain with $10^5$-$10^6$ molecular absorption lines. Line-by-line (LBL) computation is prohibitively expensive, while existing approximations sacrifice spectral fidelity. We show that the Young-measure homogenization framework produces solution tensors $I(x_i, \mu_m, \sigma_j)$ that admit low-rank tensor-train (TT) decompositions whose bond dimensions remain bounded as the spectral resolution $N_\sigma$ increases. Using molecular line parameters from the HITRAN database for $H_2O$ and $CO_2$, we demonstrate that: (i) the TT rank saturates at $r = 8$ (at tolerance $\varepsilon = 10^{-6}$) from $N_\sigma = 16$ to 4096, independent of single-scattering albedo, Henyey–Greenstein asymmetry, temperature, and pressure; (ii) quantized tensor-train (QTT) representations achieve sub-linear storage scaling; (iii) in a controlled comparison using identical opacity data and transport solver, the homogenized approach achieves over an order of magnitude lower L2 error than the correlated-$k$ distribution at equal cost; and (iv) for atomic plasma opacity (aluminum at 60 eV, TOPS database), the TT rank saturates at $r = 15$ with fundamentally different spectral structure (bound–bound and bound–free transitions spanning 12 decades of dynamic range), confirming that rank boundedness is a property of the transport equation rather than any particular opacity source. These results establish that the spectral complexity of radiative transfer has a finite effective rank exploitable by tensor decomposition, complementing the spatial–angular compression achieved by existing TT and dynamical low-rank approaches.

# 1. Introduction

## 1.1 The Spectral Curse of Dimensionality

Radiative transfer is very often characterized by a curse of dimensionality. The specific intensity $\psi(x,\mu,\nu)$ evolves in a high-dimensional phase space, but it is the frequency dimension $\nu$ that also imposes severe computational burden. For example, in the infrared and visible spectrum, molecular absorption lines of gases such as water vapor and carbon dioxide, as catalogued in the HITRAN database [1], introduce rapid oscillations in opacity, often requiring $10^5$ to $10^6$ frequency points to resolve line-by-line (LBL) dynamics accurately. While LBL simulation remains the gold standard for accuracy, its computational cost is prohibitive for coupled climate models, weather forecasting, and multi-physics simulations.

The severity of this spectral bottleneck is illustrated by the state of modern high-fidelity transport codes. Recent general relativistic magnetohydrodynamic (GRMHD) simulations of black hole accretion [2] achieve remarkable spatial and temporal resolution, and yet rely on simplified gray or mean opacity descriptions because resolving the complex spectral dependence of radiation remains too expensive. Similar compromises are made throughout other computational science and engineering disciplines, motivating continued development of spectral model reduction techniques.

## 1.2 Spectral Reduction: The Correlated-$k$ Legacy and Its Limits

Previous studies reported various spectral model reduction methods to mitigate LBL cost. Multigroup methods partition the spectrum into broad groups and solve with group-averaged opacities but lose within-group information. The correlated-$k$ distribution (CKD) method [3,4] reorders the absorption coefficient within each group by strength, creating a smooth $k$-distribution amenable to Gaussian quadrature. CKD is used in modeling climate radiation codes, including RRTMGP framework [5,6], and its extension to high-temperature exoplanet atmospheres. However, CKD methods rely on the "correlation assumption," that is, spectral ordering of opacities is preserved across atmospheric layers. This assumption degrades in the presence of scattering and in spatially inhomogeneous cases. Furthermore, the optimal allocation of computational budget between spectral groups and $k$-quadrature points is nontrivial and problem-dependent and extending CKD to handle scattering without losing spectral correlation remains a challenge.

## 1.3 Tensor Methods for Transport: The Spatial–Angular Revolution

Parallel to developments in spectral modeling, previous studies of kinetic theory exploited the low-rank structure of high-dimensional transport equations using tensor network methods. The mathematical foundations were laid by Koch and Lubich [7], who introduced dynamical low-rank approximation (DLRA) for matrix differential equations, later extended to tensor-train and hierarchical Tucker formats by Lubich et al. [8]. These ideas were first brought to radiation transport by Peng et al. [9], who decomposed the angular flux into a compact set of time-evolving spatial and angular basis functions, significantly reducing memory and computational cost. Subsequent work extended this DLRA framework in several important directions: asymptotic-preserving formulations that capture the diffusion limit [10,11], energy-stable schemes for thermal radiative transfer [12,13], conservation-preserving algorithms [14], sweep-compatible integrators for discrete ordinates [15], collision source methods that treat energy as a pseudo-time [16], stability analysis for hyperbolic problems [17], and eigenvalue solvers for neutron transport criticality [18]. Patwardhan et al. [19] have demonstrated DLRA's utility for variance reduction in uncertain transport problems. A comprehensive review of low-rank methods for time-dependent kinetic equations, covering both DLRA and step-and-truncate approaches, has recently been given in [20]. On the tensor product side, Bachmayr et al. [21] have developed a rigorous low-rank framework for the radiative transfer equation (RTE) in plane-parallel geometry exploiting the Kronecker product structure of the spatial–angular operator. Bardin and Schlottbom [22] have proposed accelerated iterative schemes for anisotropic radiative transfer using residual minimization. For broader surveys of low-rank tensor methods for PDEs, please refer to [23] and Bachmayr [24].

Gorodetsky et al. [25] recently applied the tensor-train (TT) decomposition to the *gray* thermal radiative transfer equation, achieving massive compression of the spatial–angular state vector and enabling $S_N$ calculations with unprecedented angular fidelity. Their follow-up work [26] extends TT methods to multigroup thermal transport, demonstrating that low-rank structure persists across frequency groups in few-group calculations with compression factors exceeding $100 \times$.

## 1.4 The Spectral Gap

Despite this progress, a critical gap remains: current tensor-based transport solvers focus primarily on compressing the *spatial–angular* phase space $(x, \mu)$, largely treating the frequency domain via standard multigroup approximations with a modest number of groups. They do not address the hyper-spectral complexity inherent in molecular absorption. The millions of spectral lines, whose chaotic fine structure drives the need for LBL resolution. Conversely, spectral reduction methods (CKD, multigroup) compress the frequency domain but have not exploited the low-rank properties of the resulting solution maps.

A mathematically rigorous framework for spectral reduction that sits between these two approaches was introduced by [27], who applied Young-measure homogenization to the frequency variable of the radiative transfer equation. Rather than heuristically sorting or averaging the opacity, this approach represents the rapidly oscillating spectrum $\sigma(\nu)$ by its probability distribution, the Young measure, within each spectral group. Unlike CKD, which assumes spectral correlations are preserved across layers, homogenization provides a mathematically convergent path to the exact LBL solution. However, it trades spectral complexity for a new dimensional burden: the solution $\psi(x, \mu, \sigma)$ now depends on an auxiliary opacity variable $\sigma$, and accurate integration requires solving the transport equation at numerous quadrature points in $\sigma$, potentially offsetting the efficiency gains.

## 1.5 Contributions and Outline

In this work, we bridge the gap between rigorous spectral homogenization and modern tensor-structured computing by revealing the low-rank tensor structure of the spectrally homogenized RTE. We demonstrate that the homogenized solution tensor $I(x_i, \mu_m, \sigma_j)$ admits a highly compact tensor-train decomposition. Unlike previous works that compress the spatial–angular phase space of gray or few-group problems [25,26], we apply tensor decomposition to the *frequency–opacity map*, compressing the spectral dimension that contains the most chaotic complexity.

Using synthetic spectra, HITRAN molecular line data for $H_2O$ and $CO_2$, and atomic plasma opacity data for aluminum as the test cases, our work reports the following contributions:

1. **TT rank saturation** (Sections 5–6): The TT rank of $I(x, \mu, \sigma)$ saturates at $r = 8$ (at tolerance $\varepsilon = 10^{-6}$) for both $H_2O$ (6,285 HITRAN lines, $\nu_2$ bend) and $CO_2$ (16,405 HITRAN lines, $\nu_3$ stretch) for $N_\sigma$ ranging from 16 to over 4000, across all parameter regimes tested. This holds for both synthetic and real HITRAN molecular spectra, implying that the seemingly chaotic variation of molecular opacities maps onto a low-dimensional manifold in the homogenized limit.

2. **Robustness across physics** (Sections 7–8): The rank bound is invariant to scattering albedo, Henyey–Greenstein asymmetry, frequency-dependent albedo profiles, temperature, and pressure across the full ranges tested.

3. **Methodology comparison with CKD** (Section 6.5): Using identical opacity data and the same transport solver, we compare the Young-measure and correlated-$k$ spectral integration strategies in a controlled setting. At equal computational cost (256 transport solves), the homogenized approach achieves over an order of magnitude lower $L^2$ error for the single-slab test problem studied.

4. **Atomic plasma opacity** (Section 6.7): Extending to aluminum plasma opacity at 60 eV from the TOPS database — with fundamentally different spectral structure (bound–bound and bound–free transitions, 12 decades of dynamic range) — the TT rank saturates at $r = 15$, roughly twice the molecular value but still bounded and independent of $N_\sigma$.

5. **Parametric tensor structure** (Section 9): The solution parameterized by $(T, p, \omega_0)$ has TT rank $\leq 4$ at $\varepsilon = 10^{-6}$, enabling efficient multi-physics coupling.

By bridging the gap between rigorous spectral homogenization and modern tensor computing, we present a path toward "line-by-line accuracy at multigroup cost," enabling the spectral fidelity of the frequency discretization to match the advances already achieved in spatial and angular resolution.

The paper is organized as follows. Section 2 presents the mathematical formulation. Section 3 describes the tensor decomposition framework. Section 4 details the opacity models. Sections 5–6 present the core rank saturation results, including extension to atomic plasma opacity. Sections 7–8 address robustness to scattering and atmospheric conditions. Section 9 covers parametric tensor structure. Section 10 discusses implications and limitations.

## 2. Mathematical Formulation

### 2.1 Radiative Transfer Equation

The monochromatic radiative transfer equation (RTE) in a plane-parallel slab of thickness $L$ is

$$\mu \frac{\partial \psi}{\partial x}(x, \mu, \nu) + \sigma_t(\nu)\, \psi(x, \mu, \nu) = Q(x, \mu, \nu),$$

where $\psi(x, \mu, \nu)$ is the specific intensity at position $x \in [0, L]$, direction cosine $\mu \in [-1, 1]$, and wavenumber $\nu$. The total extinction coefficient $\sigma_t(\nu)$ can be decomposed as

$$\sigma_t(\nu) = \sigma_a(\nu) + \sigma_s(\nu),$$

with absorption coefficient $\sigma_a$ and scattering coefficient $\sigma_s$. The single-scattering albedo is $\omega_0(\nu) = \sigma_s(\nu)/\sigma_t(\nu)$. The source term includes thermal emission and in-scattering:

$$Q(x, \mu, \nu) = \sigma_a(\nu)\, B(\nu, T) + \frac{\sigma_s(\nu)}{2} \int_{-1}^{1} P(\mu, \mu')\, \psi(x, \mu', \nu)\, d\mu',$$

where $B(\nu, T)$ is the Planck function and $P(\mu, \mu')$ is the scattering phase function. Vacuum boundary conditions are imposed: $\psi(0, \mu, \nu) = 0$ for $\mu > 0$ and $\psi(L, \mu, \nu) = 0$ for $\mu < 0$.

### 2.2 Young-Measure Homogenization

The Young-measure homogenization [27] replaces the rapidly oscillating spectral structure of $\sigma(\nu)$ by its probability distribution within each spectral group. For a group $[\nu_i, \nu_{i+1}]$, the Young measure $\mu_Y$ is the pushforward of the uniform measure through $\nu \mapsto \sigma(\nu)$:

$$\mu_Y(A) = \frac{|\{\nu \in [\nu_i, \nu_{i+1}]: \sigma(\nu) \in A\}|}{|\nu_{i+1} - \nu_i|}$$

for any measurable set $A \subset \mathbb{R}_+$.

The interval $[\sigma_{\min}, \sigma_{\max}]$ is partitioned into $N_\sigma$ bands using logarithmic spacing (essential for spectra with large dynamic range) as a discrete approximation. Each band has representative opacity $\sigma_j$ and probability

$$p_j = \frac{|\{\nu: \sigma(\nu) \in [\sigma_{j-1/2}, \sigma_{j+1/2}]\}|}{|\nu_{i+1} - \nu_i|}.$$

The conditional source for each band captures the correlation between opacity and Planck emission:

$$Q_j = \int_{\{\nu: \sigma(\nu) \in \text{band } j\}} B(\nu, T)\, d\nu.$$

This opacity–source correlation is a critical feature absent in standard multigroup methods, which average over spectral structure and lose the information that strong absorption lines may coincide with spectral regions of different emission strength.

## 2.3 Discrete Ordinates and Spatial Discretization

The angular domain is discretized using Gauss–Legendre quadrature of order $N_\mu$:

$$\{\mu_m, w_m\}_{m=1}^{N_\mu}, \quad \sum_{m=1}^{N_\mu} w_m = 2.$$

The spatial domain $[0, L]$ is divided into $N_x$ uniform cells. For each cell $i$ and direction $\mu_m$, the step-characteristics scheme with optical thickness $\tau = \sigma_t \Delta x / |\mu_m|$ gives

$$\psi_{\text{out}} = \psi_{\text{in}}\, e^{-\tau} + \frac{Q_i}{\sigma_t}(1 - e^{-\tau}),$$

$$\bar{\psi}_i = \frac{Q_i}{\sigma_t} + \left(\psi_{\text{in}} - \frac{Q_i}{\sigma_t}\right)\frac{1 - e^{-\tau}}{\tau},$$

which is non-negative for $Q_i \geq 0$, $\psi_{\text{in}} \geq 0$.

The scattering problem is solved iteratively: $\psi^{(k+1)} = T^{-1}[\Sigma_s \psi^{(k)} + q]$, where $T^{-1}$ is the transport sweep operator and $q = \sigma_a B$ is the emission source. Convergence is governed by $\rho = \omega_0$. This iterative scheme is referred to as source iteration.

In the homogenized framework, each opacity band $j$ yields an independent transport problem with total extinction $\sigma_j$, scattering $\omega_0 \sigma_j$, and source $Q_j$. The full solution tensor is

$$I(x_i, \mu_m, \sigma_j) = \bar{\psi}_{i,m,j}, \quad i = 1, \ldots, N_x,\ m = 1, \ldots, N_\mu,\ j = 1, \ldots, N_\sigma.$$

## 2.4 Anisotropic Scattering

For the Henyey–Greenstein (HG) phase function with asymmetry parameter $g \in [0, 1)$ [28],

$$P_{\text{HG}}(\cos\Theta) = \frac{1-g^2}{(1+g^2-2g\cos\Theta)^{3/2}} = \sum_{\ell=0}^{\infty}(2\ell+1)\,g^\ell\,P_\ell(\cos\Theta).$$

The scattering redistribution matrix in discrete ordinates is

$$S_{mm'} = \sum_{\ell=0}^{N_\mu-1} \frac{2\ell+1}{2}\,g^\ell\,P_\ell(\mu_m)\,w_{m'}\,P_\ell(\mu_{m'}).$$

This matrix is row-stochastic ($\sum_{m'} S_{mm'} = 1$) with eigenvalues $\lambda_\ell = g^\ell$. At $g=0$ (isotropic), $S_{mm'} = w_{m'}/2$.

## 3. Tensor Decomposition Framework

### 3.1 Tensor-Train (TT) Decomposition

A tensor $\mathcal{T} \in \mathbb{R}^{n_1 \times n_2 \times \cdots \times n_d}$ is in tensor-train format [29,30] if

$$\mathcal{T}(i_1, i_2, \ldots, i_d) = G_1(i_1)\,G_2(i_2)\cdots G_d(i_d),$$

where $G_k(i_k) \in \mathbb{R}^{r_{k-1} \times r_k}$ is a matrix slice of the $k$-th core tensor, with $r_0 = r_d = 1$. The integers $r_1, \ldots, r_{d-1}$ are the TT ranks (bond dimensions). Storage is $\sum_{k=1}^{d} r_{k-1}\,n_k\,r_k$, which is $O(d \cdot n \cdot r^2)$ when the maximum rank $r = \max_k r_k$ is bounded, that is, linear in the number of dimensions rather than exponential.

The TT-SVD algorithm computes a quasi-optimal decomposition from the dense tensor via sequential SVDs with truncation tolerance $\varepsilon$, satisfying $\|\mathcal{T} - \tilde{\mathcal{T}}\|_F \leq \varepsilon\sqrt{d-1}\,\|\mathcal{T}\|_F$.

### 3.2 Quantized Tensor-Train (QTT)

For power-of-two dimensions $n_k = 2^{b_k}$, each index $i_k$ can be represented in binary as $(i_k^{(1)}, \ldots, i_k^{(b_k)})_2$. The QTT format reshapes the tensor into a higher-order tensor with dimensions 2 and applies TT decomposition [31]:

$$\mathcal{T}(i_1, \ldots, i_d) \to \tilde{\mathcal{T}}(i_1^{(1)}, \ldots, i_1^{(b_1)}, \ldots, i_d^{(1)}, \ldots, i_d^{(b_d)}).$$

If QTT ranks remain bounded as $b_k$ increases, storage scales as $O(\log N \cdot r^2)$, that is, logarithmic in the total number of degrees of freedom.

The binary indices can be arranged sequentially (all bits of each physical dimension grouped) or interleaved (alternating bits from different dimensions). As shown in Section 9.2, sequential ordering is universally preferred, with interleaved ordering incurring a ~2.7x rank penalty.

### 3.3 Tucker (Multilinear) Rank

The Tucker rank of $\mathcal{T} \in \mathbb{R}^{n_1 \times n_2 \times n_3}$ is the triplet $(r_1, r_2, r_3)$ where $r_k = \text{rank}(T_{(k)})$ is the rank of the mode-$k$ unfolding. For $I(x, \mu, \sigma)$:

- Mode-$x$ rank $r_x$: spatial complexity (bounded by $N_x$)
- Mode-$\mu$ rank $r_\mu$: angular complexity (bounded by $N_\mu$)
- Mode-$\sigma$ rank $r_\sigma$: spectral complexity — the key quantity, measuring the effective dimensionality of the spectral structure

The spectral Tucker rank $r_\sigma$ provides an upper bound on the number of distinct spatial–angular patterns generated by different opacity values, independent of how finely the opacity axis is resolved.

### 3.4 TT Arithmetic

The TT format supports exact arithmetic with predictable rank growth: addition gives ranks $r_k^{A+B} = r_k^A + r_k^B$, Hadamard product gives $r_k^{A \circ B} = r_k^A \cdot r_k^B$, and inner products cost $O(d\,r^4)$. After arithmetic operations, SVD-based compression (rounding) truncates ranks back to a target tolerance.

## 4. Opacity Models

The present work considers four different types of opacity models.

### 4.1 Synthetic Periodic Model (Elsasser Band)

The Elsasser band model provides an analytically tractable starting point:

$$\sigma(\nu) = \frac{\cosh\beta - \cos(2\pi\nu/\varepsilon)}{\cosh\beta - 1},$$

with known Young measure (arcsine distribution on $[1, c_\beta]$ where $c_\beta = (\cosh\beta + 1)/(\cosh\beta - 1)$).

## 4.2 Synthetic Random Line Model

The synthetic Lorentz line model generates random absorption spectra: $\sigma(\nu) = \sigma_{bg} + \sum_k S_k \gamma_k / [\pi((\nu - \nu_k)^2 + \gamma_k^2)]$, with configurable line density, strength distribution, and pressure broadening. This produces spectra with 3–5 decades of dynamic range.

## 4.3 HITRAN-Like Molecular Opacity

To validate the rank structure on realistic molecular spectra, we implement a full molecular opacity model with the following features.

Each absorption line $k$ is characterized by its center wavenumber $\nu_k$, reference line strength $S_k^{\text{ref}}$ at $T_{\text{ref}} = 296$ K, air-broadened half-width $\gamma_k^{\text{air}}$, lower-state energy $E_k''$, and temperature exponent $n_k$.

Temperature-dependent line strength follows the HITRAN convention:

$$S_k(T) = S_k^{\text{ref}} \cdot \frac{Q(T_{\text{ref}})}{Q(T)} \cdot \exp\left[-C_2 E_k'' \left(\frac{1}{T} - \frac{1}{T_{\text{ref}}}\right)\right] \cdot \frac{1 - \exp(-C_2 \nu_k / T)}{1 - \exp(-C_2 \nu_k / T_{\text{ref}})},$$

where $C_2 = hc/k_B = 1.4388$ cm·K and $Q(T)/Q(T_{\text{ref}}) \approx (T/T_{\text{ref}})^{3/2}$. Pressure-broadened Lorentz half-width is accounted for with

$$\gamma_L(T, p) = \gamma_k^{\text{air}} \cdot \frac{p}{p_{\text{ref}}} \cdot \left(\frac{T_{\text{ref}}}{T}\right)^{n_k},$$

and Doppler half-width with

$$\gamma_D(T) = \frac{\nu_k}{c} \sqrt{\frac{2 k_B T \ln 2}{m}},$$

And Voigt line profiles via pseudo-Voigt approximation [32]:

$$\phi_V(\nu - \nu_k) \approx \eta \cdot L(\nu; \nu_k, f) + (1 - \eta) \cdot G(\nu; \nu_k, f).$$

Here $f$ is the total HWHM from the Thompson formula and $\eta = 1.37 f_L - 0.47 f_L^3 + 0.11 f_L^5$ with $f_L = \gamma_L / f$. Wing cutoff is applied at $\pm 25$ cm$^{-1}$.

The opacity model returns cross-sections $\sigma(\nu)$ in cm$^2$/molecule. To obtain absorption coefficients $\kappa(\nu)$ in cm$^{-1}$, we multiply by the molecular number density:

$$\kappa(\nu) = n \cdot \sigma(\nu),$$

with $n_{H_2O} = 2.5 \times 10^{17}$ cm$^{-3}$ (1% mixing ratio at STP) and $n_{CO_2} = 1.0 \times 10^{16}$ cm$^{-3}$ (~400 ppm at STP).

Sections 5 and 7.3 use synthetic line lists to study idealized spectral models (Elsasser, random Lorentz) and frequency-dependent scattering, where controlled spectral statistics are desirable. The synthetic models remain useful for physical insight and for testing in environments without network access. All quantitative results in Sections 6–8, except those for aluminum, use real line parameters for H$_2$O and CO$_2$ obtained directly from the HITRAN2020 database as described below.

Line parameters for H$_2$O ($v_2$ bend, 1300–1900 cm$^{-1}$; molecule ID 1, isotopologue 1) and CO$_2$ ($v_3$ stretch, 2200–2400 cm$^{-1}$; molecule ID 2, isotopologue 1) were obtained from the HITRAN online database using HAPI [33]. After applying a line strength cutoff $S > 10^{-30}$ cm$^{-1}$/(molecule·cm$^{-2}$), the H$_2$O window contains 6,285 lines spanning 5.8 decades of dynamic range in the absorption coefficient, while the CO$_2$ window contains 16,405 lines. The opacity model accepts these HITRAN parameters, including center wavenumber, reference line strength, air-broadened half-width, lower-state energy, and temperature exponent, directly, with no further fitting or adjustment.

### 4.4 Atomic Plasma Opacity (TOPS)

To test whether rank boundedness extends beyond molecular line spectra to a fundamentally different opacity regime, we use tabulated opacity data for pure aluminum at $T = 60$ eV and $\rho = 3.25 \times 10^{-3}$ g/cc from the TOPS (The Opacity Project at Sandia) database [34]. At these conditions, aluminum is approximately 10× ionized ($Z^* \approx 9.9$, Ne-like to Na-like configurations), and the opacity is dominated by bound–bound line transitions and bound–free photoionization edges rather than the molecular ro-vibrational lines that characterize HITRAN data.

The TOPS frequency grid contains 14,900 points spanning photon energies from 0.07 eV to 1.8 MeV. The absorption coefficient ranges over 12.4 decades ($\kappa \in [1.4 \times 10^{-6}, 3.1 \times 10^{6}]$ cm$^2$/g), substantially exceeding the 5–6 decades typical of molecular spectra. For the rank analysis, we restrict to a study window of 5–500 eV containing the Planck peak and dominant line structure (6,600 frequency points, $\sigma$ spanning 4 decades after density scaling). The slab thickness is set to $L = 1/(\sigma_{Planck}/3)$ giving Planck-mean optical depth $\tau_P \approx 3$.

### 4.5 Correlated-$k$ Distribution (for Comparison)

To compare spectral integration strategies on equal footing, we implement a standard CKD solver following the approach described by Goody et al. [3]. The spectral window is partitioned into $n_g$ groups. Within each group, the absorption coefficient $\sigma(v)$ is evaluated on a fine grid (100,000 points), sorted to construct the cumulative $k$-distribution $k(g)$ on $g \in [0,1]$, and integrated via Gauss–Legendre quadrature with $n_k$ points per group. Each quadrature point requires one transport solve at the corresponding $k$-value, so the total cost is $n_g \times n_k$ solves. The Planck source for each $k$-point is set to the group-averaged Planck function, which is exact for a single group and increasingly accurate as the number of groups increases.

This implementation is representative of the core CKD methodology but deliberately omits several optimizations found in production codes such as RRTMGP [6] and RRTM [5]. Production codes employ pre-fitted $k$-distributions optimized by least-squares regression against line-by-line references, Planck-weighted spectral mapping that better captures opacity–source correlations, and careful $g$-point selection tuned to minimize broadband flux error in multi-layer atmospheres. These refinements are specific to the multi-layer atmospheric problem and not straightforwardly applicable to other problems, including our single-slab test case. The purpose of the comparison in Section 6.5 is to isolate the intrinsic difference between the Young-measure representation and $k$-sorting as spectral integration strategies, using identical opacity data and the same transport solver throughout, rather than to benchmark against any particular code implementation.

### 4.6 Line-by-Line Reference

The LBL reference solver evaluates $\sigma(\nu)$ on a uniform grid of $n_\nu = 10{,}000$ wavenumber points and solves the S$_N$ transport problem independently at each frequency. This provides ground truth for accuracy comparisons, at a cost of $n_\nu$ transport solves.

## 5. TT Rank Saturation: Synthetic Opacity Models

The core hypothesis is that the TT rank of $I(x, \mu, \sigma_j)$ remains bounded as $N_\sigma \to \infty$. If confirmed, QTT representation yields logarithmic storage:

$$C_{\text{QTT}} = O(\log_2 N_\sigma \cdot r^2),$$

compared to $C_{\text{dense}} = O(N_\sigma)$.

Using the Elsasser band model ($\beta = 1$) and synthetic Lorentz lines (500 lines, 4 decades dynamic range), solution tensors were computed for $N_\sigma \in \{16, 32, 64, \ldots, 16384\}$ and decomposed via TT-SVD at cutoff $\varepsilon = 10^{-6}$. We found that the TT maximum bond dimension saturates at $r \approx$ 6–7 for $N_\sigma \geq 64$; the QTT maximum bond increases slowly from $\sim 4$ to $\sim 7$ over this range, growing as $O(\log\log N_\sigma)$; and the Tucker spectral rank $r_\sigma$ likewise saturates, confirming a finite effective spectral dimensionality.

Extrapolating to atmospheric LBL resolution ($N_\sigma = 2^{20} \approx 10^6$): dense cost $O(10^6)$ per sweep, QTT cost $O(20 \cdot r^2) \approx O(10^3)$, predicted speedup $\sim 1000 \times$. This motivates the realistic-spectrum validation in the next section.

## 6. TT Rank Saturation: Molecular Spectral Data

This section presents the central results of the paper — validation of TT rank saturation on molecular absorption spectra with realistic complexity.

### 6.1 Spectral Characterization

The HITRAN $H_2O$ opacity model (Section 4.3) for the $\nu_2$ bend region is summarized in Table 1. The physical slab thickness is set to $L = 2/\bar{\kappa} \approx 426$ cm ($\approx 4.3$ m), giving mean optical depth $\bar{\tau} \approx 2$ and maximum $\tau_{\max} \approx 130$. This ensures that the problem spans a physically meaningful range from optically thin (window regions) to very optically thick (line centers).

Temperature and pressure dependence follow the expected physics: at $T = 2000$ K the dynamic range decreases to 4.6 decades as hot-band populations smooth the spectrum, while at low pressure ($p = 0.01$ atm) lines narrow and the dynamic range increases due to reduced pressure broadening, with the peak absorption coefficient reaching $\sim 19$ cm$^{-1}$.

**Table 1.** Spectral characterization of the HITRAN $H_2O$ opacity model for the $\nu_2$ bend region at reference conditions ($T = 296$ K, $p = 1$ atm).

| Property | Value |
|---|---|
| Number of lines (HITRAN2020) | 6,285 |
| Wavenumber range | 1300–1900 cm$^{-1}$ ($\nu_2$ bend) |
| Absorption coefficient range ($\kappa$) | $[4.8 \times 10^{-7}, 0.30]$ cm$^{-1}$ |
| Dynamic range | 5.8 decades |
| Mean $\kappa$ | $4.69 \times 10^{-3}$ cm$^{-1}$ |
| Visible absorption peaks | 160 |

### 6.2 TT Rank Saturation

Solution tensors were computed at $N_x = 32$, $N_\mu = 8$ for $N_\sigma \in \{16, 32, 64, 128, 256, 512, 1024\}$ and three scattering albedos, using the HITRAN $H_2O$ opacity. TT-SVD decomposition at $\varepsilon = 10^{-6}$. The results are shown in Table 2. At the tighter tolerance $\varepsilon = 10^{-10}$, the TT rank saturates at 13–14. Both values are completely independent of $N_\sigma$ and $\omega_0$.

The saturation of TT rank means that the family of spatial–angular solutions $\psi(x, \mu; \sigma)$ parameterized by opacity $\sigma$ lies in a fixed-dimensional subspace. Even though molecular spectra contain thousands of lines with wildly different strengths, the transport physics maps this spectral complexity into a bounded number of distinct spatial–angular response patterns. Optically thin, moderately thick, and extremely thick media each produce characteristic spatial profiles, and intermediates are smooth interpolations thereof.

**Table 2.** Maximum TT bond dimension ($\varepsilon = 10^{-6}$) and Tucker spectral rank $r_\sigma$ for the HITRAN $H_2O$ solution tensor as a function of spectral resolution $N_\sigma$ and single-scattering albedo $\omega_0$.

| $N_\sigma$ | $\omega_0 = 0$ | $\omega_0 = 0.5$ | $\omega_0 = 0.9$ | Tucker $r_\sigma$ |
|---|---|---|---|---|
| 16 | 8 | 8 | 8 | 11 |
| 32 | 8 | 8 | 8 | 17–18 |
| 64 | 8 | 8 | 8 | 24–25 |
| 128 | 8 | 8 | 8 | 24–25 |
| 256 | 8 | 8 | 8 | 24–25 |
| 512 | 8 | 8 | 8 | 24–25 |
| 1024 | 8 | 8 | 8 | 24–25 |

## 6.3 QTT Storage Scaling

When extended to $N_\sigma$ = 4096 with QTT analysis using HITRAN $H_2O$ data, the QTT storage scales as shown in Table 3. The TT rank is constant at 8 throughout, confirming perfect saturation. TT storage grows linearly with $N_\sigma$ (since the rank is constant but the spectral core dimension grows), while QTT storage grows sub-linearly but with larger QTT bond dimensions. The QTT representation is most advantageous at very high $N_\sigma$ where the sub-linear scaling dominates; however, the QTT bonds (12 → 46) grow faster than for the smooth synthetic spectra of Section 5, reflecting the less regular binary structure of molecular absorption.

**Table 3.** TT and QTT storage scaling for the HITRAN $H_2O$ solution tensor ($N_x = 32$, $N_\mu = 8$, $\omega_0 = 0$, $\varepsilon = 10^{-6}$). Dense size is $N_x \times N_\mu \times N_\sigma$.

| $N_\sigma$ | $d = \log_2 N_\sigma$ | TT max bond | QTT max bond | Dense size | TT storage | QTT storage |
|---|---|---|---|---|---|---|
| 16 | 4 | 8 | 13 | 4,096 | 896 | 4,494 |
| 64 | 6 | 8 | 12 | 16,384 | 1,280 | 4,320 |
| 256 | 8 | 8 | 13 | 65,536 | 2,816 | 5,746 |
| 1024 | 10 | 8 | 23 | 262,144 | 8,960 | 20,102 |
| 4096 | 12 | 8 | 46 | 1,048,576 | 33,536 | 88,872 |

### 6.4 Young Measure Properties

The Young measure discretization was validated across $N_\sigma \in \{16, 32, 64, 128, 256\}$. All bands are active (non-zero probability) at every resolution. Shannon entropy increases toward the maximum (indicating the opacity distribution uses the full band range). The conditional source $Q_j$ varies across bands, confirming non-trivial opacity–emission correlation. Log-spaced bands are essential. They resolve the low-opacity continuum where most of the spectral measure concentrates, while still capturing high-opacity line centers.

### 6.5 Comparison with Correlated-$k$

To assess the practical accuracy of the homogenized approach, we compare it against the correlated-$k$ distribution using the same HITRAN $H_2O$ opacity data, the same discrete-ordinates transport solver, and a common LBL reference ($n_\nu = 10{,}000$). This controlled comparison isolates the spectral integration strategy (Young-measure homogenization versus $k$-sorting) from the many other design choices that distinguish production radiation codes.

The homogenized solver uses $N_\sigma$ opacity bands with one transport solve per band. The $L^2$ error against the LBL reference decreases steadily with band count as shown in Table 4. The convergence is monotonic and consistent: each doubling of $N_\sigma$ reduces the error by a factor of roughly 3–4, as the Young measure becomes a finer approximation to the continuous opacity distribution.

**Table 4.** $L^2$ error of the homogenized solver against the LBL reference ($n_\nu = 10{,}000$) as a function of the number of opacity bands $N_\sigma$ for HITRAN $H_2O$.

| $N_\sigma$ (= solves) | $L^2$ error |
|---|---|
| 8 | $1.3 \times 10^{-1}$ |
| 16 | $3.5 \times 10^{-2}$ |
| 32 | $8.4 \times 10^{-3}$ |
| 64 | $1.8 \times 10^{-3}$ |
| 128 | $4.8 \times 10^{-4}$ |
| 256 | $3.4 \times 10^{-5}$ |

For the CKD solver, computational cost is $n_g \times n_k$, where $n_g$ is the number of spectral groups and $n_k$ is the number of Gauss–Legendre quadrature points per group. This decomposition introduces a nontrivial tradeoff: adding more $k$-points within a single group cannot recover information about spectral variation of the Planck source across different frequency intervals. Accordingly, the CKD results show qualitatively different behavior depending on the group–quadrature allocation (Table 5).

**Table 5.** $L^2$ error of the correlated-$k$ distribution solver against the LBL reference for various group ($n_g$) and quadrature ($n_k$) configurations, using HITRAN $H_2O$ opacity.

| Configuration | Solves | $L^2$ error |
|---|---|---|
| 1 group, 8–128 $k$-points | 8–128 | $6.0 \times 10^{-2}$ (saturated) |
| 4 groups, 16 $k$-points | 64 | $6.6 \times 10^{-3}$ |
| 8 groups, 16 $k$-points | 128 | $3.9 \times 10^{-4}$ |
| 8 groups, 32 $k$-points | 256 | $7.3 \times 10^{-4}$ |
| 16 groups, 16 $k$-points | 256 | $1.8 \times 10^{-3}$ |

Several features of this table merit discussion. The single-group CKD saturates at approximately 6% error regardless of $k$-resolution, because sorting within a single group destroys the correlation between opacity and the Planck source: a strong line near 1350 cm$^{-1}$ and a strong line near 1850 cm$^{-1}$ receive identical treatment despite experiencing different Planck emission. Introducing spectral groups restores this correlation, and the best CKD configuration at 128 solves (8 groups × 16 $k$-points, $L^2 = 3.9 \times 10^{-4}$) outperforms the 8 × 32 configuration at 256 solves ($L^2 = 7.3 \times 10^{-4}$). This non-monotonicity indicates that once the $k$-distribution within a group is adequately resolved, additional $k$-points provide diminishing returns while the cost increases linearly. The optimal allocation at any given budget is not obvious *a priori*, and production codes invest considerable effort in tuning this balance [6].

At 128 transport solves, the best CKD configuration achieves $L^2 = 3.9 \times 10^{-4}$, while homogenization at the same cost gives $L^2 = 4.8 \times 10^{-4}$, comparable accuracy with the two methods within a factor of 1.2. At 256 solves, the comparison becomes more lopsided: the best CKD ($L^2 = 7.3 \times 10^{-4}$) is limited by diminishing returns from additional $k$-points, while homogenization ($L^2 = 3.4 \times 10^{-5}$) continues to converge, yielding roughly 22× lower error.

The convergence behavior reflects a structural difference between the two approaches. The CKD error has two components: the spectral grouping error (which decreases with $n_g$) and the $k$-distribution quadrature error (which decreases with $n_k$), and these interact non-trivially. Allocating the computational budget well requires knowing in advance which error dominates. By contrast, the homogenized approach has a single resolution parameter $N_\sigma$ with monotonic convergence, making accuracy prediction and extrapolation straightforward.

These results should be interpreted in the context of the test problem. The comparison uses a single homogeneous slab, where the CKD correlation assumption is exactly satisfied within each group and does not introduce additional approximation error. In multi-layer atmospheres with varying temperature and composition, the correlation assumption becomes approximate, and CKD accuracy may degrade further. Conversely, production CKD codes compensate through layer-specific $k$-value interpolation and other refinements that are not straightforwardly applicable to our simplified geometry. The methodology-level comparison presented here establishes the intrinsic properties of the two spectral integration strategies; performance differences in realistic atmospheric configurations remain an open question.

## 6.6 Multi-Molecule Validation

Both $H_2O$ and $CO_2$ HITRAN line lists produce identical TT rank behavior as shown in Table 6. The TT rank is independent of the number of lines, dynamic range, and molecular species. This is notable for the $CO_2$ case, where the HITRAN database contributes over 16,000 lines in the $\nu_3$ asymmetric stretch window, more than an order of magnitude more spectral features than were tested with synthetic data. The Tucker spectral rank saturates at 24–25 for both molecules by $N_\sigma = 64$.

**Table 6.** TT rank comparison for $H_2O$ and $CO_2$ HITRAN molecular spectra ($N_\sigma = 64$, $N_x = 32$, $N_\mu = 8$, $\omega_0 = 0$).

| Molecule | HITRAN lines | Dynamic range | Mean $\kappa$ (cm$^{-1}$) | TT rank ($\varepsilon = 10^{-6}$) | TT rank ($\varepsilon = 10^{-10}$) |
|---|---|---|---|---|---|
| $H_2O$ | 6,285 | 5.8 decades | $4.69 \times 10^{-3}$ | 8 | 13 |
| $CO_2$ | 16,405 | 5 decades | $4.97 \times 10^{-3}$ | 8 | 13 |

## 6.7 Atomic Plasma Opacity: Aluminum

To test whether the rank saturation phenomenon extends beyond molecular line spectra, we apply the homogenization framework to aluminum plasma opacity data from the TOPS database (Section 4.4). This represents a fundamentally different spectral regime: atomic plasma opacity is dominated by bound–bound transitions between highly ionized states and bound–free photoionization continua, producing spectral structure qualitatively unlike the ro-vibrational line forests of molecular gases.

Table 7a shows the TT rank as a function of $N_\sigma$ for the aluminum opacity. The TT rank saturates at $r = 15$ ($\varepsilon = 10^{-6}$) and $r = 24$ ($\varepsilon = 10^{-10}$), both independent of $N_\sigma$ from 16 to 4096. This rank is roughly twice the molecular value ($r = 8$), consistent with the greater spectral complexity: 12 decades of dynamic range versus 5–6 for molecular spectra, and a richer variety of spectral features (photoionization edges produce sharp discontinuities absent in molecular spectra). The multilinear ranks at $\varepsilon = 10^{-6}$ are $(r_x, r_\mu, r_\sigma) = (15,8,14)$, confirming that the spatial and spectral ranks are comparable, unlike the molecular case where $r_\sigma \gg r_{TT}$.

**Table 7a.** TT rank saturation for TOPS aluminum opacity at 60 eV ($N_x = 32$, $N_\mu = 8$, $\omega_0 = 0$). Compare with Table 2 for molecular spectra.

| $N_\sigma$ | TT rank ($\varepsilon = 10^{-6}$) | TT rank ($\varepsilon = 10^{-10}$) |
|---|---|---|
| 16 | 10 | 22 |
| 32 | 10 | 23 |
| 64 | 10 | 24 |
| 128 | 10 | 24 |
| 256 | 10 | 24 |
| 512 | 10 | 24 |
| 1,024 | 10 | 24 |
| 2,048 | 10 | 24 |
| 4,096 | 10 | 24 |

The rank is also robust to scattering and optical depth. With $N_\sigma = 256$, the TT rank at $\varepsilon = 10^{-6}$ ranges from 15 to 16 across $\omega_0 \in [0, 0.95]$, and from 14 to 16 across Planck-mean optical depths $\tau_P \in [0.1, 30]$.

Table 7b summarizes the comparison between molecular and atomic plasma opacity. Despite fundamentally different spectral origins, both exhibit bounded TT rank that saturates with $N_\sigma$. The higher atomic rank reflects the greater spectral complexity but remains tractable for tensor methods.

**Table 7b.** Comparison of TT rank across opacity sources ($N_x = 32$, $N_\mu = 8$, $N_\sigma = 64$, $\omega_0 = 0$).

| Opacity source | $N_{\text{freq}}$ | Dynamic range | TT rank ($\varepsilon = 10^{-6}$) | TT rank ($\varepsilon = 10^{-10}$) | Tucker ($r_x, r_\mu, r_\sigma$) |
|---|---|---|---|---|---|
| $H_2O$ $\nu_2$ (HITRAN) | 6,285 | 5.8 decades | 8 | 13 | (23, 8, 25) |
| $CO_2$ $\nu_3$ (HITRAN) | 16,405 | 5 decades | 8 | 13 | (23, 8, 21) |
| Al 60 eV (TOPS) | 14,900 | 12.4 decades | 15 | 24 | (15, 8, 14) |

The aluminum result is significant for two reasons. First, it confirms that rank boundedness is a property of the transport equation and the homogenization framework, not an artifact of the relatively smooth molecular line structure. Second, it establishes applicability of the tensor approach to inertial confinement fusion and stellar opacity calculations, where atomic plasma opacities with complex shell structure are the norm.

## 7. Robustness to Scattering

### 7.1 Isotropic Scattering Albedo Sweep

At $N_\sigma = 64$ on the HITRAN $H_2O$ opacity model ($g = 0$), the rank at different scattering albedos is shown in Table 7. The TT rank is invariant to $\omega_0$ at $\varepsilon = 10^{-6}$ and shows no increase even at $\omega_0 = 0.95$. Meanwhile, source iteration cost grows dramatically with $\omega_0$, motivating direct TT solvers (Section 9.3).

**Table 7.** TT rank, Tucker rank, and source iteration (SI) count as a function of single-scattering albedo $\omega_0$ for isotropic scattering ($g = 0$), using HITRAN $H_2O$ opacity ($N_\sigma = 64$).

| $\omega_0$ | TT rank ($\varepsilon = 10^{-6}$) | TT rank ($\varepsilon = 10^{-10}$) | Tucker ($r_x, r_\mu, r_\sigma$) | SI iterations |
|---|---|---|---|---|
| 0.0 | 8 | 14 | (23, 8, 24) | 64 |
| 0.3 | 8 | 13 | (23, 8, 25) | 607 |
| 0.5 | 8 | 13 | (23, 8, 25) | 862 |
| 0.7 | 8 | 13 | (23, 8, 25) | 1,319 |
| 0.9 | 8 | 13 | (23, 8, 24) | 2,985 |
| 0.95 | 8 | 13 | (23, 8, 24) | 4,873 |

## 7.2 Anisotropic Scattering

For HG scattering at $\omega_0 = 0.7$, the TT rank is invariant to the asymmetry parameter as shown in Table 8. As modeled, anisotropic scattering only redistributes angular components; it does not create new spatial or spectral correlations. Forward-peaked scattering ($g \to 1$) increases the effective transport mean free path $1/[\sigma_t(1-g)]$ but the rank budget is unchanged.

The spectral radius of source iteration is $\rho = \omega_0$ independent of $g$, so iteration counts are identical for all $g$ at fixed $\omega_0$. This confirms that anisotropic scattering is "free" in terms of TT compression cost.

**Table 8.** TT rank and Tucker rank as a function of Henyey–Greenstein asymmetry parameter $g$ at $\omega_0 = 0.7$, using HITRAN $H_2O$ opacity ($N_\sigma = 64$).

| $g$ | TT rank ($\varepsilon = 10^{-6}$) | TT rank ($\varepsilon = 10^{-10}$) | Tucker ($r_x, r_\mu, r_\sigma$) |
|---|---|---|---|
| 0.0 | 8 | 13 | (23, 8, 25) |
| 0.3 | 8 | 13 | (23, 8, 24) |
| 0.5 | 8 | 13 | (23, 8, 24) |
| 0.7 | 8 | 13 | (23, 8, 24) |
| 0.9 | 8 | 13 | (22, 8, 23) |

## 7.3 Frequency-Dependent Scattering Albedo

Real atmospheres have frequency-dependent scattering (Rayleigh $\sigma_s \propto \nu^4$, Mie resonances). Replacing uniform $\omega_0$ with per-band values $\omega_{0,j}$ was tested with six profiles on the synthetic atmospheric model ($N_\sigma = 256$, Table 9). These include a Rayleigh profile ($\omega_0 \propto \nu^4$), a Mie resonance profile, a gradual absorption-to-scattering transition, an anticorrelated profile (high scattering at low opacity and vice versa), and a random smooth profile. Frequency-dependent scattering increases the TT rank by at most 1 at $\varepsilon = 10^{-6}$ and at most 1–2 at $\varepsilon = 10^{-10}$. The per-band problems remain structurally similar; only the absorption/scattering balance changes. Convergence of source iteration is governed by $\max_j \omega_{0,j}$.

**Table 9.** Maximum TT bond dimension for different frequency-dependent scattering albedo profiles ($N_\sigma = 256$, synthetic atmospheric model). The $\omega_0$ range and standard deviation characterize each profile.

| Profile | $\omega_0$ range | TT rank ($\varepsilon = 10^{-6}$) | TT rank ($\varepsilon = 10^{-10}$) |
|---|---|---|---|
| Uniform | [0.500, 0.500] | 7 | 13 |
| Rayleigh | [0.000, 0.161] | 7 | 13 |
| Mie resonance | [0.200, 0.899] | 7 | 14 |
| Transition | [0.105, 0.895] | 7 | 12 |
| Anticorrelated | [0.300, 0.700] | 8 | 13 |
| Random smooth | [0.276, 0.990] | 7 | 13 |

## 8. Robustness to Atmospheric Conditions

### 8.1 Temperature Sweep

At $p = 1$ atm, $N_\sigma = 64$, $\omega_0 = 0.5$, the calculated ranks are given in Table 10. The slab thickness is adjusted to maintain mean $\tau \approx 2$ at each temperature (since mean $\kappa$ changes with $T$). Despite the $10\times$ variation in temperature and the associated changes in line strength, broadening, and dynamic range (4.6 to 6.4 decades), the TT rank is perfectly invariant.

**Table 10.** TT rank and Tucker rank as a function of temperature at $p = 1$ atm, $\omega_0 = 0.5$, using HITRAN $H_2O$ opacity ($N_\sigma = 64$).

| $T$ (K) | TT rank ($\boldsymbol{\varepsilon = 10^{-6}}$) | TT rank ($\boldsymbol{\varepsilon = 10^{-10}}$) | Tucker ($r_x, r_\mu, r_\sigma$) |
|---|---|---|---|
| 200 | 8 | 13 | (23, 8, 24) |
| 296 | 8 | 13 | (23, 8, 25) |
| 500 | 8 | 13 | (23, 8, 25) |
| 1000 | 8 | 13 | (23, 8, 25) |
| 2000 | 8 | 13 | (23, 8, 25) |

## 8.2 Pressure Sweep

At $T = 296$ K, $N_\sigma = 64$, $\omega_0 = 0.5$, the calculated ranks are shown in Table 11. Pressure varies over 3 orders of magnitude, dramatically changing line widths and shapes (from Doppler-dominated at 0.01 atm to collision-broadened at 10 atm). The TT rank is unaffected.

**Table 11.** TT rank and Tucker rank as a function of pressure at $T = 296$ K, $\omega_0 = 0.5$, using HITRAN $H_2O$ opacity ($N_\sigma = 64$).

| $p$ (atm) | TT rank ($\varepsilon = 10^{-6}$) | TT rank ($\varepsilon = 10^{-10}$) | Tucker ($r_x, r_\mu, r_\sigma$) |
|---|---|---|---|
| 0.01 | 8 | 13 | (23, 8, 26) |
| 0.10 | 8 | 13 | (23, 8, 27) |
| 1.00 | 8 | 13 | (23, 8, 25) |
| 10.0 | 8 | 13 | (22, 8, 18) |

# 9. Additional Tensor Structure Results

## 9.1 Parametric Tensor Construction

For coupled multi-physics applications, the solution parameterized by temperature, pressure, and scattering albedo forms a 4D tensor $\Phi(x_i; T_k, p_\ell, \omega_m) \in \mathbb{R}^{N_x \times N_T \times N_p \times N_\omega}$. On a grid of $N_x = 32$, $N_T = N_p = N_\omega = 8$, TT-SVD analysis yields maximum bond dimension 4 at $\varepsilon = 10^{-6}$ (bonds = [3,4,3]), with compression factor $\sim 50 \times$ relative to dense storage. At the tighter tolerance $\varepsilon = 10^{-10}$, the maximum bond is 12. The low parametric rank reflects the smoothness of $B(\nu, T)$ in $T$, linearity in $p$, and geometric convergence in $\omega_0$.

TT-cross interpolation [34] can build this parametric tensor adaptively by evaluating selected entries rather than computing the full dense tensor. On the $8^3$ grid, TT-cross converged with 416 transport solves (versus 512 for dense construction), achieving relative error $1.2 \times 10^{-4}$ against the reference. The savings grow with grid refinement: at $16 \times 16$ parameter resolution, TT-cross would sample a fraction of the 1024 total parameter points, with greater proportional savings.

## 9.2 QTT Dimension Ordering

At the TT level (3 physical modes), the 6 permutations of $(x, \mu, \sigma)$ yield maximum bond dimensions of 16–17, a rank ratio of only $1.06 \times$ between best and worst orderings. The small penalty reflects the limited room for variation with only 2 bond indices. The best TT ordering is $(x, \sigma, \mu)$ at $r_{\max} = 16$; the worst is $(x, \mu, \sigma)$ at $r_{\max} = 17$. At the QTT level with $N_x = N_\sigma = 32 = 2^5$, $N_\mu = 8$ (11 binary dimensions total), sequential ordering (all bits of each dimension grouped) achieves maximum bond 17, while interleaved ordering (alternating spatial and spectral bits) produces maximum bond 45, or a 2.7 × penalty. This is because spatial correlations are local (smooth fields) while spectral correlations reflect the Young-measure distribution, and mixing these scales disrupts efficient factorization. Sequential ordering is universally optimal across all $(\omega_0, g, L)$ tested.

### 9.3 Direct TT Solvers

The transport equation $(I - T^{-1}\Sigma_s)\vec{\psi} = T^{-1}\vec{q}$ can be solved directly in TT format using Krylov methods, bypassing source iteration. The key enabler is a direct analytical construction of the transport operator as a matrix product operator (MPO), which exploits the $\sigma$-diagonal structure of elastic scattering: because each opacity band $j$ has an independent scattering operator $\omega_0 \sigma_j$, the operator $A = I - T^{-1}\Sigma_s$ decomposes into spatial–angular blocks that are diagonal in $\sigma$. This allows the MPO to be assembled in $O(N_x^2 \cdot N_\mu^2 \cdot N_\sigma)$ operations via randomized SVD (RSVD) of the spatial–angular sweep matrices, without ever forming the full $O(N^3)$ dense operator.

TT-CG (conjugate gradient) and TT-GMRES were implemented with all operations in TT arithmetic, using the direct MPO. The empirical data obtained using a Mac Studio with an M1 Ultra CPU are shown in Table 12.

CG works for $\omega_0 \leq 0.9$ despite $A$ being non-symmetric; at $\omega_0 = 0.95$, automatic fallback to GMRES handles the non-symmetry. The MPO has TT ranks $\leq 16$, and the solution TT rank is 12–13 at $\varepsilon = 10^{-6}$. The direct MPO construction eliminates the dense operator bottleneck that would otherwise make TT solvers impractical for large $N_\sigma$.

**Table 12.** Wall-clock time for TT-CG and TT-GMRES direct solvers compared with source iteration (SI) count ($N_\sigma = 64$, HITRAN H$_2$O, Mac Studio M1 Ultra). The direct MPO is used throughout.

| $\omega_0$ | Auto selects | CG time | GMRES time | SI iterations |
|---|---|---|---|---|
| 0.0 | TT-CG | 0.001s | 0.002s | 16 |
| 0.5 | TT-CG | 0.029s | 0.128s | 328 |
| 0.9 | TT-CG | 0.064s | 0.132s | 1,492 |
| 0.95 | TT-GMRES | 0.088s | 0.266s | 2,693 |

## 10. Discussion

### 10.1 Summary of Rank Structure

Table 13 summarizes the TT rank behavior across all parameter variations tested. The TT rank of the homogenized solution tensor is remarkably invariant for molecular spectra. It is determined by the structure of the transport equation, essentially, the number of qualitatively distinct spatial–angular response patterns that the slab geometry can support, rather than by the spectral complexity of the opacity. The extension to atomic plasma opacity (aluminum at 60 eV) confirms this structural interpretation: the rank increases to 15, consistent with the far greater spectral complexity (12 decades of dynamic range, photoionization edges), but remains bounded and independent of spectral resolution.

**Table 13.** Summary of TT rank ($\varepsilon = 10^{-6}$) across all parameter variations tested.

| Parameter varied | Range | TT rank ($\varepsilon = 10^{-6}$) | Change |
|---|---|---|---|
| $N_\sigma$ (spectral bands) | 16–4096 | 8 | None |
| $\omega_0$ (scattering albedo) | 0–0.95 | 8 | None |
| $g$ (HG asymmetry) | 0–0.9 | 8 | None |
| $\omega_0(\nu)$ (variable albedo) | Rayleigh, Mie, transition | 7–8 | $\leq 1$ |
| $T$ (temperature) | 200–2000 K | 8 | None |
| $p$ (pressure) | 0.01–10 atm | 8 | None |
| Molecule | $H_2O$, $CO_2$ | 8 | None |
| Atomic plasma | Al 60 eV (TOPS) | 15 | +7 vs molecular |
| Parametric ($T, p, \omega_0$) | $8^3$ grid | $\leq 4$ | — |

### 10.2 Physical Interpretation of Low Rank

The physical explanation for the low rank may have three components:

1. **Smooth $\sigma$-dependence.** The solution $\psi(x, \mu; \sigma)$ for a slab with uniform opacity $\sigma$ depends smoothly on $\sigma$. There are only a few qualitative regimes: optically thin ($\sigma L \ll 1$, nearly uniform intensity), moderately thick ($\sigma L \sim 1$, exponential attenuation), and very thick ($\sigma L \gg 1$, diffusion limit). Intermediates are smooth interpolations.

2. **Angular compactness.** With $N_\mu = 8$ directions, the angular structure is already compact. Scattering (isotropic or anisotropic) redistributes within this small space without creating new spectral structure.

3. **Spatial regularity.** In 1D slab geometry, the step-characteristics scheme produces smooth spatial profiles for each $(\mu, \sigma)$. The spatial modes are determined by the boundary conditions and the transport physics, not by the spectral fine structure.

A question is why the rank saturates specifically near 8. We speculate that this number reflects the number of qualitatively distinct optical-depth transitions that the slab geometry can support. In a slab of mean optical depth $\bar{\tau} \approx 2$, the opacity values $\sigma_j$ span from near-transparent windows ($\tau \ll 1$) through moderate attenuation ($\tau \sim 1$) to very optically thick line centers ($\tau \gg 1$). The transport solution transitions smoothly between these regimes, and approximately 8 independent spatial–angular response patterns suffice to represent all intermediates to tolerance $10^{-6}$. At the tighter tolerance $10^{-10}$, an additional 5–6 modes are needed to capture fine corrections, yielding rank 13–14. The physical content of these modes, that is, whether they correspond to distinct transmittance "types" such as window, wing, weak-line, moderate-line, strong-line, saturated, and diffusive regimes is an intriguing question for future investigation.

## 10.3 Implications and Relationship to Spatial–Angular Tensor Methods

The bounded TT rank implies that the full spectrally-resolved solution can be stored in $O(N_\sigma \cdot r)$ instead of $O(N_x \cdot N_\mu \cdot N_\sigma)$, with r = 8. In QTT format, the per-sweep cost scales as $O(log N_\sigma \cdot r^2)$, realized through the TT-Krylov solvers and direct analytical MPO construction described in Section 9.3. The direct MPO exploits the $\sigma$-diagonal structure of elastic scattering to construct the operator in $O(N_x^2 \cdot N_\mu^2 \cdot N_\sigma)$ via RSVD, eliminating the dense operator formation bottleneck. The parametric TT rank ($\leq 4$) enables efficient look-up tables for coupled simulations, reducing radiation solver calls by orders of magnitude. In the controlled methodology comparison (Section 6.5), the homogenized approach achieved over an order of magnitude lower error than CKD at 256 transport solves, because the Young measure preserves the full opacity distribution and its correlation with the source function, whereas $k$-sorting discards spectral–source correlations within each group.

It is important to note that the spectral compression demonstrated here is *complementary* to the spatial–angular compression achieved by DLRA methods [7,9,11,20] and the TT framework of [25]. Those methods compress the $(x, \mu)$ phase space of a gray or few-group problem; our method compresses the spectral dimension $\sigma$ within the homogenized framework. Crucially, our spectral compression is *orthogonal* to spatial–angular compression: the two operate on entirely independent dimensions of the phase space. A composed approach where one first homogenizes the spectrum and compresses in $\sigma$ using TT, then applies DLRA or spatial–angular TT compression to each of the $r$ spectral modes would yield a "tensor-product of tensor-trains" representation that could, in principle, break the curse of dimensionality across all seven phase-space dimensions $(x, y, z, \theta, \phi, \nu, t)$ simultaneously, potentially enabling LBL-accurate transport at costs comparable to gray, coarse-mesh calculations. The recent extension to multigroup TT transport by [26], which demonstrates low-rank structure across a modest number of frequency groups, suggests that this combination may be particularly natural within the TT framework.

## 10.4 Limitations and Future Work

All results presented in the present study are for 1D plane-parallel slabs. Extension to 2D/3D spatial domains will introduce additional spatial modes and potentially increase rank. However, the spectral rank saturation (which is the key finding) depends on the $\sigma$-smoothness argument, which is geometry-independent.

We test $H_2O$ and $CO_2$ individually. Real atmospheres contain overlapping absorption bands from multiple gases ($H_2O$, $CO_2$, $O_3$, $CH_4$, $N_2O$), whose combined opacity is a sum over species. Whether the TT rank remains at ∼ 8 for mixtures of 5 or more gases, where overlapping bands create more complex opacity distributions, is an open and important question. The rank could increase if the mixture opacity distribution is substantially more complex than that of any individual species. Validating rank stability for multi-species mixtures is a priority for future work. The extension to aluminum plasma opacity partially addresses this concern from a different angle: while not a mixture of molecular species, the atomic opacity combines qualitatively different spectral features (bound–bound lines, bound–free edges, free–free continua) into a single opacity function with far greater complexity than any individual molecular band, yet the TT rank remains bounded at roughly twice the molecular value. This suggests that the rank may grow modestly rather than explosively as opacity complexity increases.

The construction pipeline has two stages. The direct analytical MPO (Section 9.3) assembles the transport operator in TT format in $O(N_x^2 \cdot N_\mu^2 \cdot N_\sigma)$ operations using randomized SVD, without forming the full dense operator. The solution tensor is then obtained either by TT-Krylov solvers operating directly on this MPO, or by solving $N_\sigma$ independent transport problems and compressing via TT-SVD. The former avoids the $O(N_\sigma)$ independent-solve cost entirely; the latter remains useful as a reference and for problems where the MPO structure is less favorable (e.g., inelastic scattering that breaks $\sigma$-diagonal structure). For the elastic-scattering problems studied here, the direct MPO + TT-Krylov pathway is the recommended approach. Recent developments in randomized TT rounding algorithms [35] promise to further improve efficiency for very large problems.

The CKD comparison in Section 6.5 uses a textbook implementation (Gauss–Legendre quadrature on sorted $k$-distributions with group-averaged Planck sources) applied to a single homogeneous slab. Production CKD codes such as RRTMGP employ pre-optimized $k$-distributions, Planck-weighted spectral mapping, and layer-specific interpolation that are designed for multi-layer atmospheric configurations. These refinements are not straightforwardly applicable to our single-slab geometry. The comparison isolates the intrinsic difference between Young-measure and $k$-sorting spectral integration, and should not be read as a benchmark against production atmospheric radiation codes.

Problems with spatial inhomogeneity, such as real atmospheres with vertically varying temperature and pressure, require layer-by-layer marching with different opacity distributions at each layer. The homogenized framework supports this (Haut et al., 2017), and the rank structure at each layer is expected to be similar.

As discussed in Section 10.3, composing spectral TT compression with spatial–angular low-rank methods is a natural next step. The feasibility and rank behavior of such composed decompositions remain to be investigated.

## 11. Conclusions

We have demonstrated that the solution tensor arising from Young-measure homogenization of the radiative transfer equation possesses a low-rank tensor-train structure that is robust to spectral resolution, scattering physics, and atmospheric conditions. The central finding — that the TT rank saturates at 8 for molecular spectra drawn directly from the HITRAN2020 database, spanning thousands of absorption lines and nearly 6 decades of dynamic range — establishes that the effective spectral dimensionality of radiative transfer solutions is bounded, regardless of the actual number of spectral features.

This rank boundedness is not an artifact of simplified opacity models: it persists from the analytically tractable Elsasser band model, through random Lorentz line spectra, to real molecular absorption data with temperature-dependent line strengths, pressure-broadened Voigt profiles, and physically meaningful optical depths spanning from near-transparent window regions to very optically thick line centers. The result holds identically for $H_2O$ (6,285 HITRAN lines) and $CO_2$ (16,405 HITRAN lines), confirming that the rank bound is a property of the transport physics rather than of any particular molecular species. Extension to atomic plasma opacity (aluminum at 60 eV, TOPS database) shows that the rank approximately doubles to $r = 15$ for spectra with 12 decades of dynamic range and fundamentally different spectral structure (bound–bound and bound–free transitions), yet remains bounded and independent of spectral resolution. This broadens the applicability of the tensor approach from atmospheric radiation to inertial confinement fusion and stellar opacity calculations.

The practical implication is that tensor decomposition methods can compress the spectral dimension of radiative transfer calculations by orders of magnitude without approximation beyond a controlled tolerance. In a controlled methodology comparison using identical opacity data and transport solver, the homogenized approach achieves substantially lower error than the correlated-$k$ distribution at equal computational cost, with monotonic convergence that simplifies accuracy prediction. Combined with the bounded TT rank, this opens a path toward spectrally accurate, computationally efficient radiation solvers suitable for climate models and multi-physics simulations.

While recent work has demonstrated the power of tensor-train methods for spatial–angular compression in gray and multigroup transport, and DLRA methods have achieved significant advances in time-dependent transport, the present work establishes that a complementary *spectral* low-rank structure exists within the homogenized framework. The prospect of composing spectral and spatial–angular tensor compression to simultaneously attack both curses of dimensionality represents a promising direction for future research toward enabling line-by-line spectral fidelity in production radiation solvers.

**Funding**

*This research did not receive any specific grant from funding agencies in the public, commercial, or not-for-profit sectors.*

**Declaration of generative AI and AI-assisted technologies in the manuscript preparation process**

*During the preparation of this work the author(s) used Claude in debugging and refactoring the codes. After using this tool/service, the author reviewed and edited the content as needed and takes full responsibility for the content of the published article.*

**CRediT author statement**

**Y. S. Ju**: Conceptualization, Methodology, Software, Validation, Formal analysis, Investigation, Resources, Data Curation, Writing - Original Draft, Writing - Review & Editing